\documentclass[aps,prd,preprint,nofootinbib]{revtex4}
\usepackage{latexsym}
\usepackage{graphics}
\usepackage{graphicx}
\usepackage{bm}
\usepackage{amsmath}
\usepackage{amsfonts}
\usepackage{amssymb}

\usepackage[figuresright]{rotating}
\begin{document}
% put your own definitions here:
%   \newcommand{\cZ}{\cal{Z}}
%   \newtheorem{def}{Definition}[section]
%new list environments to replace itemize and enumerate
%\newenvironment{Itemize}{\begin{list}{$\bullet$}%
%    {\setlength{\topsep}{0.2mm}\setlength{\partopsep}{0.2mm}%
%    \setlength{\itemsep}{0.2mm}\setlength{\parsep}{0.2mm}}}%
%    {\end{list}}
%    \newcounter{enumct}
%    \newenvironment{Enumerate}{\begin{list}{\arabic{enumct}.}%
%    {\usecounter{enumct}\setlength{\topsep}{0.2mm}%
%    \setlength{\partopsep}{0.2mm}\setlength{\itemsep}{0.2mm}%
%    \setlength{\parsep}{0.2mm}}}{\end{list}}
%   ...

\def\be{\begin{equation}}
\def\ee{\end{equation}}
\def\bea{\begin{eqnarray}}
\def\eea{\end{eqnarray}}
\newcommand{\ggam}{\mbox{$\gamma\gamma\,$}}
\newcommand{\ggww}{\mbox{$\gamma\gamma\to W^+W^-\,$}}
\newcommand{\ttbs}{\char'134}
%\newcommand{\AmS}{{\protect\the\textfont2
% A\kern-.1667em\lower.5ex\hbox{M}\kern-.125emS}}

% add words to TeX's hyphenation exception list
%\hyphenation{author another created financial paper re-commend-ed Post-Script}

% declarations for front matter
\title{ Charge asymmetries in \bf\boldmath{$\gamma\gamma \to \mu^+\mu^- + \nu_{\mu}
\overline{\nu}_\mu$ / $\gamma \gamma \to W^\pm\mu^\mp +\nu_\mu$}
with polarized  photons}

\author{{D.~A.~Anipko},$^{1}$ {M.~Cannoni},$^{2,\;3}$ {I.~F.~Ginzburg},$^{1}$ {O.~Panella}$^{3}$ and {A.~V.~Pak}$^{1}$}

\affiliation{$^{1}$ Sobolev Institute of Mathematics and Novosibirsk State University, Novosibirsk, 630090, Russia 
}

\affiliation{
$^{2}$Dipartimento di Fisica, Universit\`a degli Studi di Perugia,
Via A. Pascoli, I-06123, Perugia, Italy
}

\affiliation{
$^{3}$Istituto Nazionale di Fisica Nucleare, Sezione di Perugia,
 Via A.~Pascoli, I-06123, Perugia, Italy
}
\date{June 16, 2003}

\begin{abstract}

It is shown that the difference in the distributions of positive
($\mu^+$) and negative charged leptons ($\mu^-$) in reactions
$\gamma \gamma \to \mu^+\mu^-+\nu\bar\nu$ and $\gamma \gamma \to
W^\pm\mu^\mp +\nu(\bar\nu)$ at $\sqrt{s}>200$ GeV leads to
observable {\em charge asymmetry} of muons which is sensitive to
New Physics effects. 

\vspace{8cm}
{\it Talk presented by I.~F.~Ginzburg at Photon 2003: International Conference on the Structure and 
Interactions of the Photon and 15th International Workshop on Photon-Photon Collisions, Frascati, 
Italy, 7-11 Apr 2003. 
}

\end{abstract}

\maketitle

\section{INTRODUCTION}

The Photon Collider~\cite{Ginzburg1} option of the next generation
linear colliders (LC)~\cite{Tesla} offers the opportunity to study
with high precision the physics of gauge bosons of the Standard
Model with sensitivity to effects coming from New Physics. High
energy photons are produced by Compton back-scattering of laser
photons from high energy electron (or positron) beams: they will
not be monochromatic but will have  an energy spectrum. The high
energy part of this spectrum will mainly include photons with
definite helicity $\lambda_i\approx \pm1$.

The SM cross section of $\gamma \gamma \to W^+W^-$ at energies
greater than $200$ GeV is about $80$ pb and remains constant up to
higher energies of interest~\cite{GKPS}. This large cross section
will ensure very high event rates. When initial photons are
circularly polarized one expects different momentum distributions
between 
the positively and negatively charged leptons from the
decay of the $W$ gauge bosons which is referred to as a {\em
charge asymmetry}. A qualitative discussion of the mechanism
originating charge asymmetries is given in Sec.~\ref{Sec2}.

One expects that a study of charge asymmetries will be a sensitive
tool to study New Physics effects. Therefore, all mechanisms
leading to this asymmetry should be investigated in detail. In the
following  we study the charge asymmetry in the Standard Model
(SM) and make preliminary considerations on how these asymmetries
change due to some possible effects of New Physics.

Numerical results (including all plots) have been obtained with
the CompHEP package~\cite{Pukhov} in a version which allows to set
the initial photons in a definite helicity state and to implement
realistic photon energy spectra. The following cuts are applied
below: (1) an angular cut on the muons scattering angles given by
$\pi- \theta_0> \theta > \theta_0$, with $\theta_0=10$ mrad
(corresponding to the TESLA detector angular limitations); (2) a
cut on muons transverse momentum $p_\perp>10$ GeV, both on each
muon and on the couple of muons. These two cuts help to suppress
the background.

\noindent We discuss two effects of charge asymmetry:\\
$\bullet$ Asymmetry of $\mu^+$ and $\mu^-$ momenta in each event,
averaged over events in the process $\gamma\gamma\to
\mu^+\mu^-\nu\bar{\nu}$ where both muons are recorded.\\
$\bullet$ Difference in distributions of $\mu^+$ and $\mu^-$ in
the processes $\gamma\gamma\to W^+\mu^-\bar{\nu}$,
$\gamma\gamma\to W^-\mu^+\nu$, where we assume $W$ reconstruction
via effective mass of two jets ($W\to q\bar{q}$ mode) or its
lepton decay. Most of the numerical results presented concern this
last approach.

$\diamondsuit$ In the following we mark polarization of an
initial state, e.g.  $(+\,-)$, that means $\lambda_1=+1$,
$\lambda_2=-1$, where $\lambda_1$ and $\lambda_2$ are helicities of
photons moving in the positive and negative direction of the collision
axis ($z$ axis) respectively.

\section{DIAGRAMS AND QUALITATIVE DESCRIPTION}
\label{Sec2}

$\bullet$ In SM, at the tree level, the process $\gamma \gamma \to
\mu^+\mu^- \nu\bar\nu$ is described by 19 diagrams. We subdivide
them in 5 types, shown in Fig.~1 (a similar classification was
given also in Ref.~\cite{Boos}). The collection of diagrams within
each type is obtained from those shown in the figure with the
exchange $+ \leftrightarrow -$, $\nu\leftrightarrow \bar{\nu}$ and
permutations of photons. In the estimates of cross sections we
denote $B=Br(W\to\mu\nu)$ and $B_Z=Br(Z\to \nu\bar{\nu})$.
\begin{figure}[htb]
\begin{center}
%\vspace{-7mm}
\includegraphics[height=8cm,width=7cm]{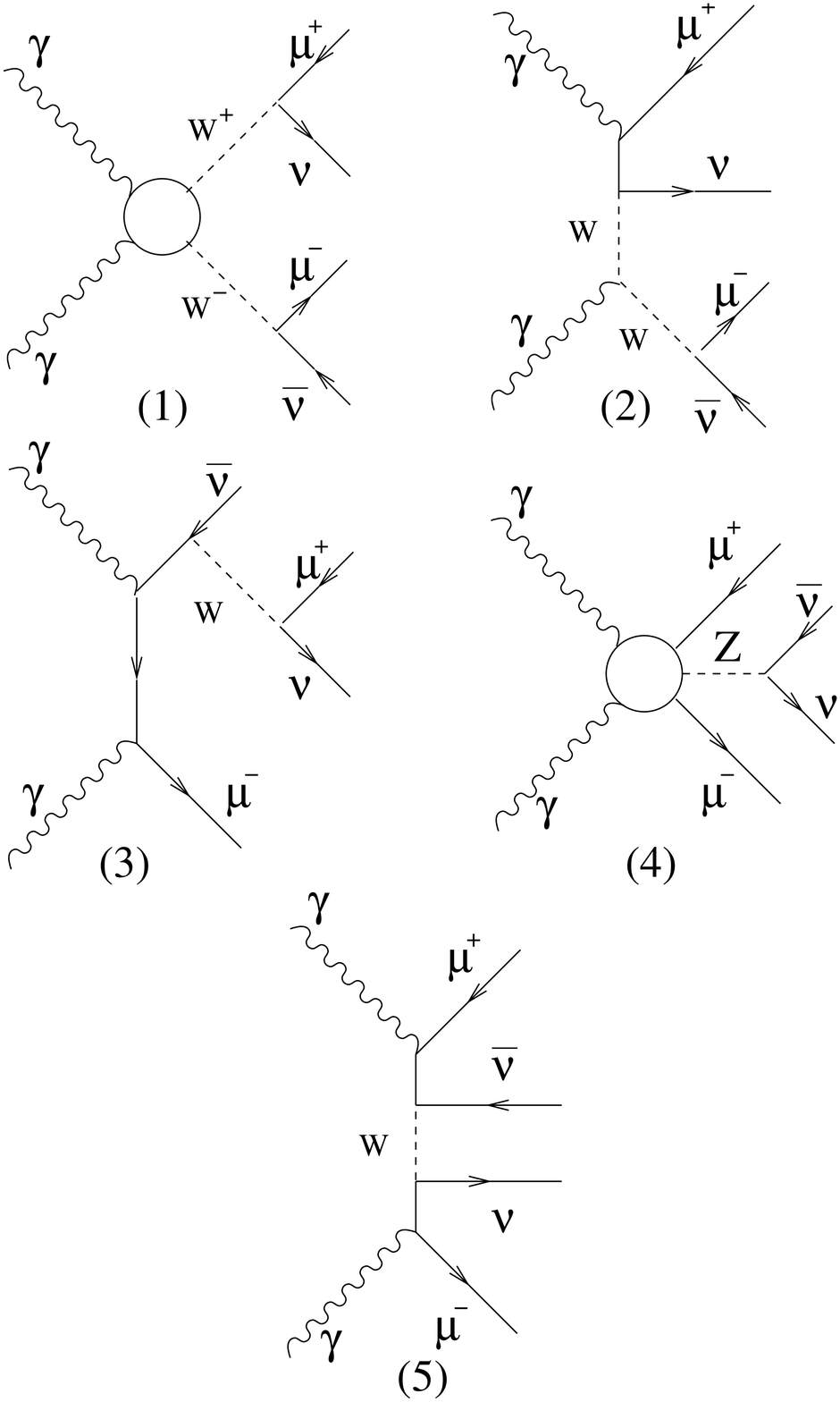}
%\vspace{-7mm} 
\caption{\sf Different classes of tree level Feynman
diagrams contributing to $\gamma \gamma \to \mu^+\mu^-
\nu\bar\nu$.} \label{diag}
\end{center}
%\vspace{-12mm}
\end{figure}
\begin{itemize}
\item[(1)]{ 3 double-resonant diagrams (DRD) of Fig.~\ref{diag}(1)
describe $WW$ production and decay. Their contribution to the
total cross section can be evaluated as $ \sigma_d \sim
(\alpha^2/M_W^2) B^2$, numerically it accounts for $\approx 70$\%
of the cross section. }
 \item[(2)]{4 single-resonant diagrams of  Fig.~\ref{diag}(2) with
$W$ exchange in $t$--channel. The corresponding contribution is
estimated as $\sigma_{sW}\sim (\alpha^3/M_W^2)B\sim
\alpha\sigma_d/B$. }
 \item[(3)]{4 single resonant diagrams with $\mu$ exchange in
$t$--channel (gauge boson bremsstrahlung), Fig.~\ref{diag}(3).
Their contribution to the total cross section is estimated as:
$\sigma_{s\mu}\sim (\alpha^3/s)B\sim \alpha\sigma_dM_W^2/(Bs)$;
numerically they provide less then 1\%.}
 \item[(4)]{6 diagrams with radiation of $Z$ boson in the process
$\gamma\gamma\to\mu^+\mu^-$, Fig.~\ref{diag}(4) give asymptotic
contribution $\sigma_Z \sim (\alpha^3/s)B_Z\sim
\alpha\sigma_dM_W^2B_Z/(B^2s)$; numerically they amount less then
1\%.}
 \item[(5)]{2  non-resonant diagrams
Fig.~\ref{diag}(5) with $\sigma_n\sim \alpha^4/M_W^2\sim
\alpha^2\sigma_dM_W^2/(B^2s)$; numerically they amount less then
1\%.}
\end{itemize}
Our analytical estimates are based on the equations for $2\to 2$
processes at $s\gg M_W^2$, assuming  for the SM gauge couplings
$g^2\sim g'^2\sim\alpha$. Numerical estimates are obtained at
$\sqrt{s}=500$ GeV with CompHEP in the Feynman gauge for the squared
contributions of each group. The value of the interference with
the dominant DRD is roughly the same, since DRD are large only in
two regions of the final phase space corresponding to $W$ resonances.
The other contributions do not have these two peaks.

The asymmetry is present in the double-resonant contribution.
There are no reasons to expect the other diagrams to contribute
to the asymmetry more than their relative contributions to the cross
section. Note that the contribution of diagrams of group 5 is at the
level of radiative corrections to the dominant contribution. The
precision provided by including this contribution is beyond the
accuracy of the tree approximation.

$\bullet$ The process $\gamma\gamma\to W^+\mu^-\bar{\nu}$ is
described by only the first 3 groups of diagrams.

{\it Therefore, in the following qualitative discussion  we  refer
to the double resonant contribution while all numerical results
and plots are obtained for the complete set of 19 diagrams.}

Let us describe the expected picture for double resonant
contribution. ({\it i}) At $\sqrt{s}>200$ GeV the \ggww\  cross
section practically does not depend on photon polarizations. ({\it
ii}) In the \ggww\ process, the  $W$ gauge bosons are distributed
mainly around the forward and backward directions with mean
transverse momentum $\sim M_W$ ($d\sigma\propto
1/(p_\bot^2+M_W^2)^2$~\cite{GKPS}). ({\it iii}) In this process
helicity conservation holds approximately (see helicity amplitudes
in~\cite{BBB}). The helicity of $W^\pm$ moving in the positive
direction of $z$ axis $\lambda_{W_1}\approx \lambda_1$, while
$\lambda_{W_2}\approx \lambda_2$, is independent on the charge
sign of $W$. ({\it iv}) Let the $z'$--axis be directed along the
$W$ momentum ($\bm{p}_W$) and $\varepsilon\approx M_W/2$ and
$p_{z'}$ are the energy and the longitudinal momentum of $\mu$ in
the $W$ rest frame. It is straightforward to show that the
distribution of muons from the decay of the $W$ gauge boson with
charge $e=\pm 1$ and helicity $\lambda=\pm 1$ in its rest frame
is
\footnote{The transverse momenta of muons are distributed
roughly isotropically relative to $W$ momentum within the interval
$p_\bot<m_W/2$.}: $\propto(\varepsilon -e\lambda p_{z'})^2$. In
other words, the distribution of muons from the decay of $W^\pm$
has a peak along $\bm{p}_W$ if $e\lambda_W=-1$ and opposite to
$\bm{p}_W$ when $e\lambda_W=+1$. These distributions are boosted
easily to  the $\gamma\gamma$ collision frame. For example when
photons have a $(-\,-)$  helicity state, produced $\mu^+$ are
distributed around the upper value of their longitudinal momentum
(both in the forward and backward direction), while produced
$\mu^-$ are concentrated near the zero value of their longitudinal
momentum. This 
%very 
boost makes the distribution in $p_\bot$ wider
in the first case and narrower in the second case.

\section{NUMERICAL RESULTS}

\begin{figure}[ht]
\begin{center}
%\vspace{-10mm}
\includegraphics[clip,height=7cm,width=7cm]{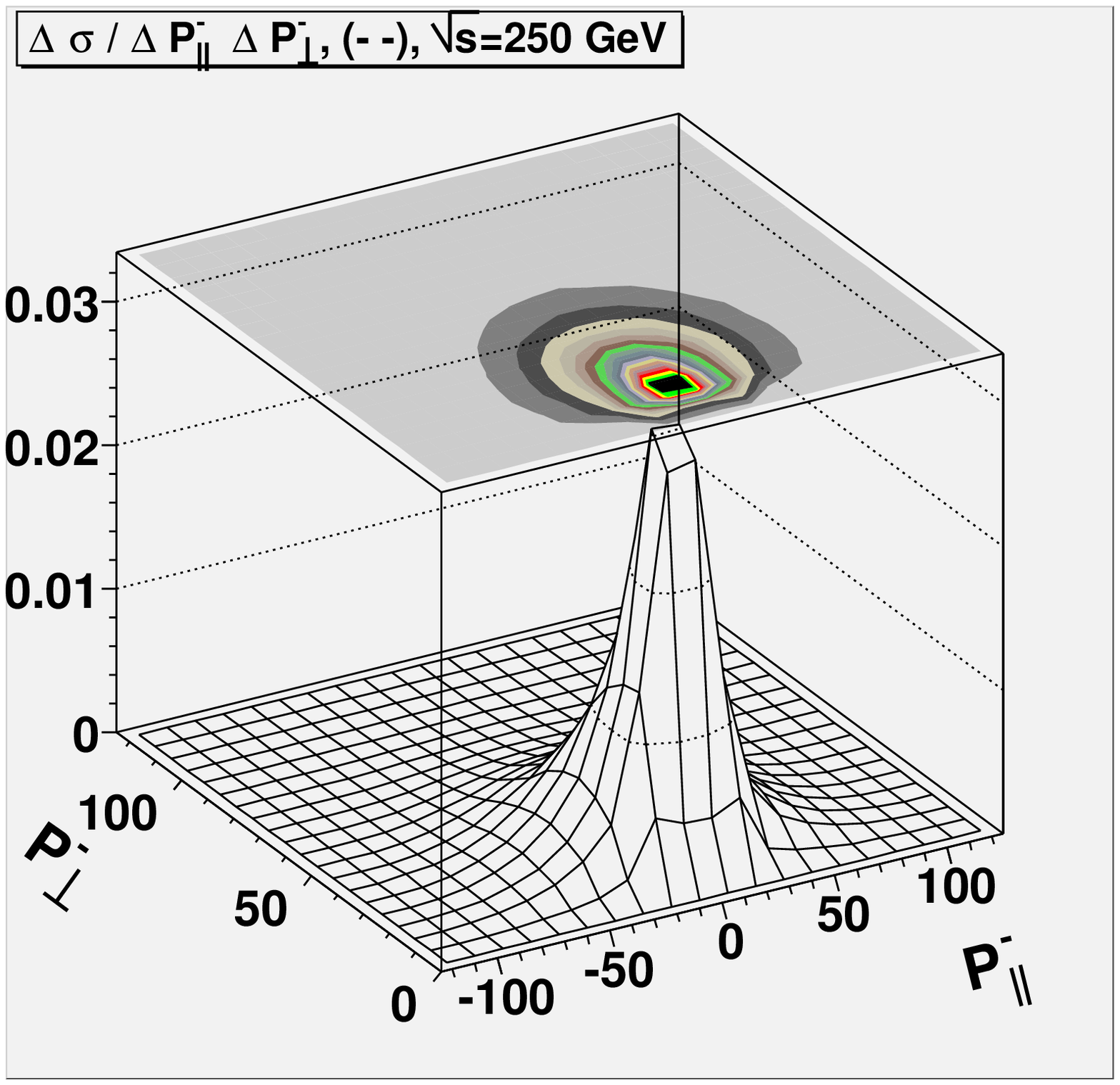}
\includegraphics[bb=0 30 473 467,clip,height=7cm,width=7cm]{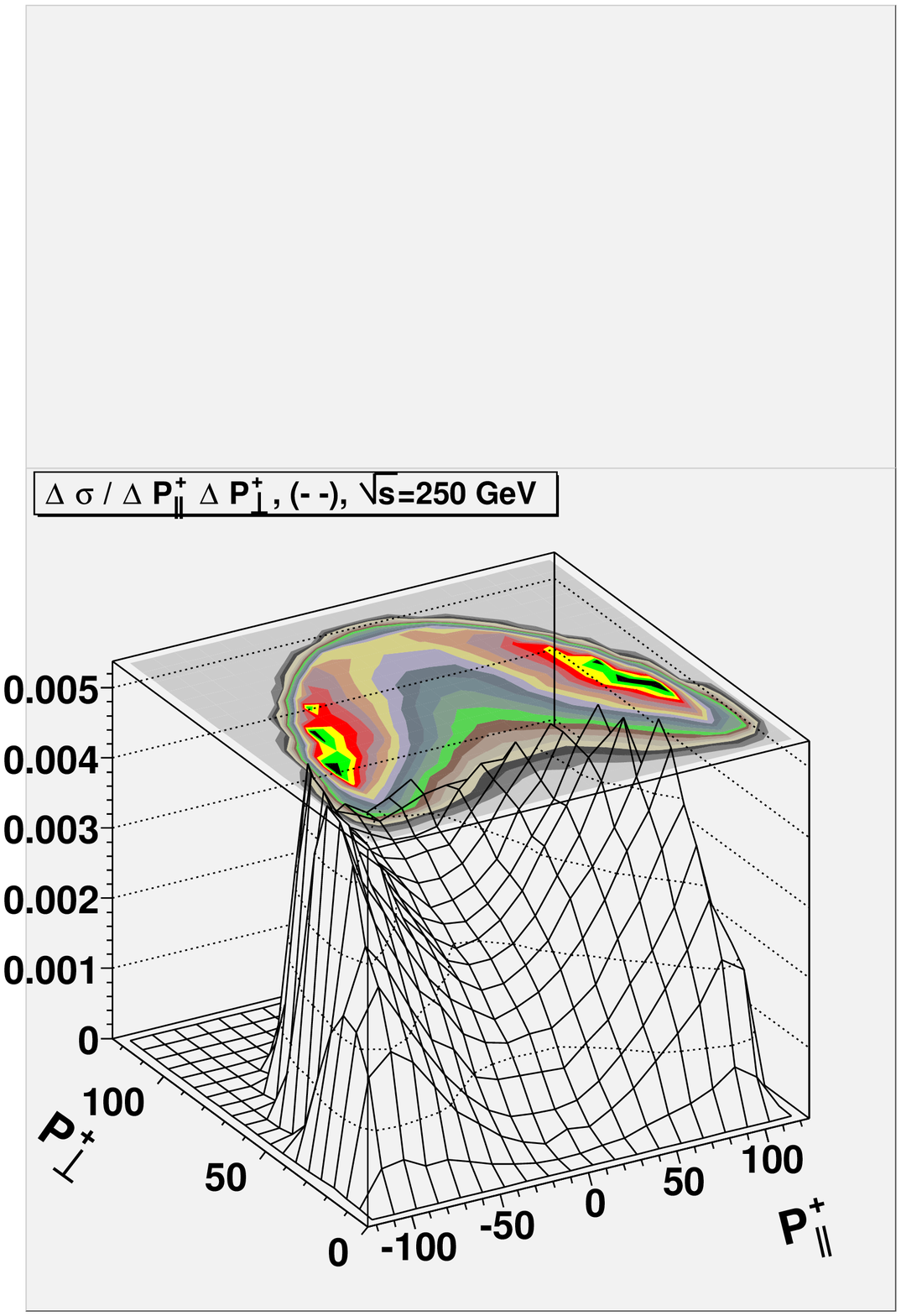}
\includegraphics[clip,height=7cm,width=7cm]{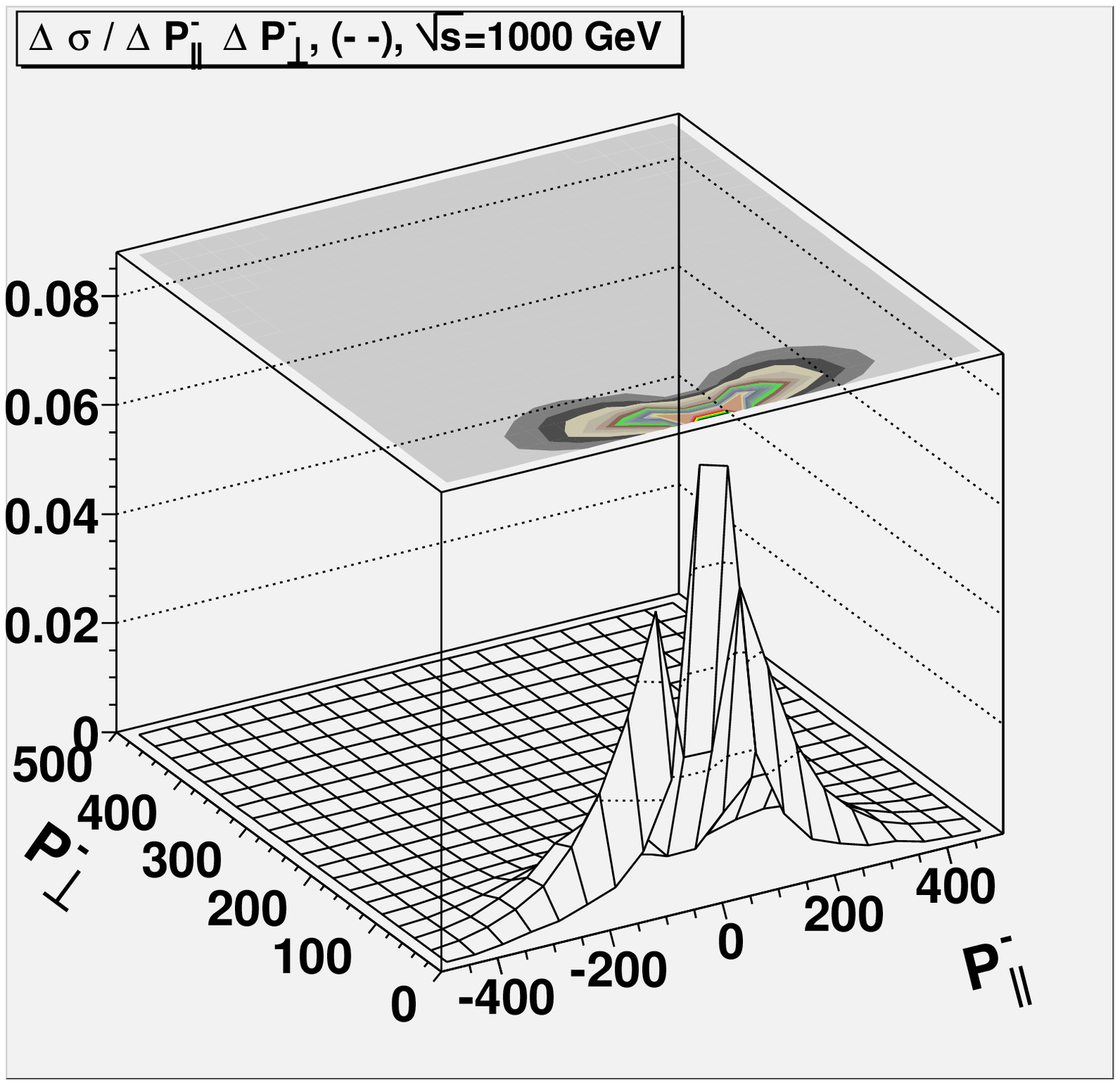}
\includegraphics[clip,height=7cm,width=7cm]{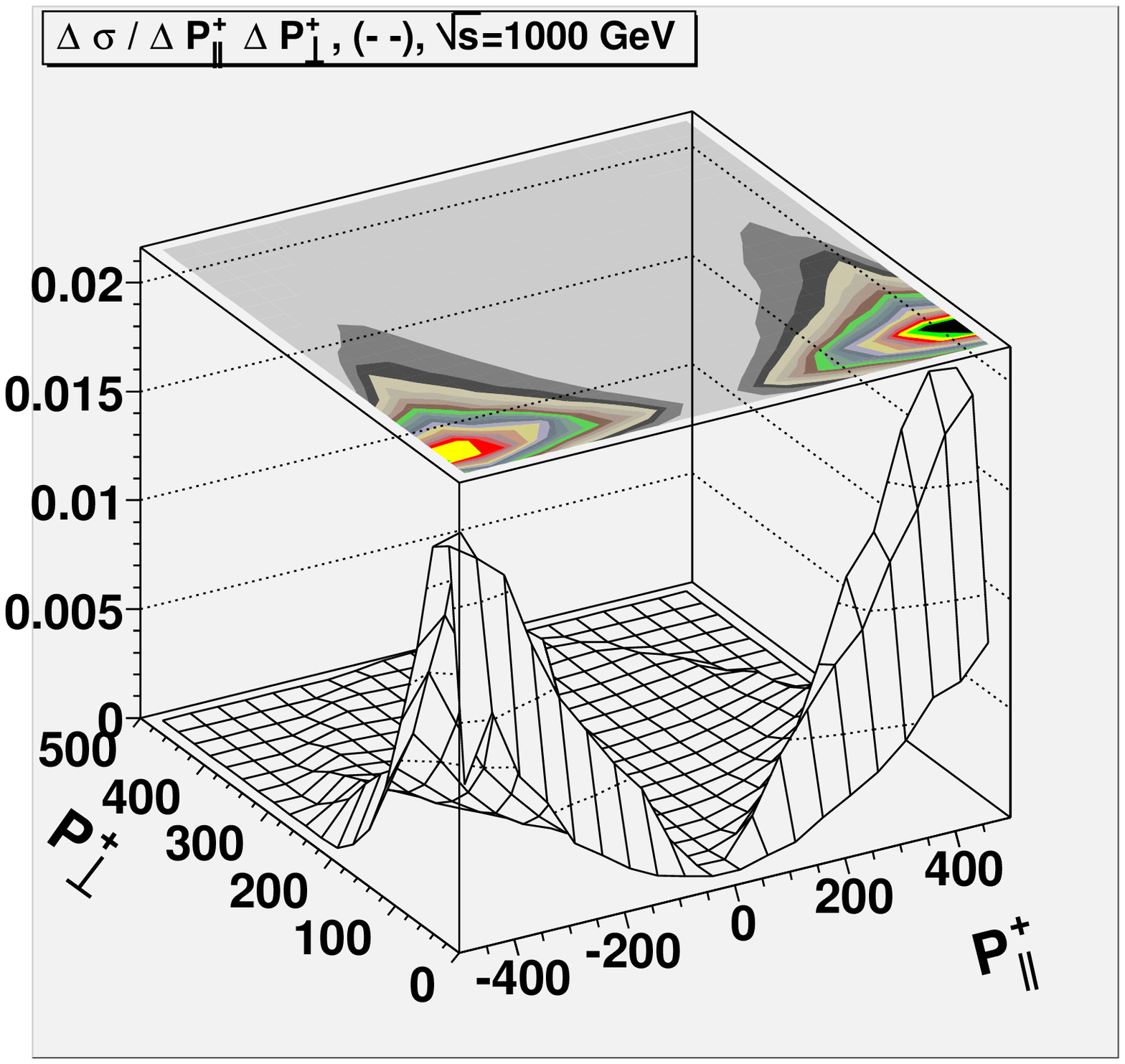}
%\vspace{-7mm}
\caption{\sf The distributions in the $(p_\parallel,p_\perp)$
plane for $(-\, -)$ helicity of colliding photons: $\mu^-$ on the
left, $\mu^+$ on the right: $\sqrt{s}=250$ GeV (top) and
$\sqrt{s}=1000$ GeV (bottom). Monochromatic beams.}
\label{energyvar}
\end{center}
%\vspace{-12mm}
\end{figure}

We present here the first results, reporting the distributions of
muons $\partial^2 \sigma/(\partial p_{\parallel}\partial
p_{\perp})$ (with $p_{\parallel}$ the component of
$\vec{p}_{\mu^\pm}$ parallel to the collision axis (taken to be
the $z$ axis) and $p_{\perp}=\sqrt{p^2_x+p^2_y}$ the transverse
momentum). Fig.~\ref{energyvar} shows such muon distributions in
the $(p_\|,\,p_\bot)$ plane for a $(-\,-)$ initial photon
polarization state with monochromatic photons at different values
of the energy. Fig.~\ref{energyvar} clearly shows the charge
asymmetry: {\em a strong difference in the distributions of
$\mu^-$ and $\mu^+$} in accordance with qualitative description
given above. The absolute value of charge asymmetry effect
decreases with energy due to the increasing importance of the
applied cuts.

{\bf\boldmath In all figures and tables below we consider 
$\sqrt{s_{\ggam}}=500$ GeV for the monochromatic case and
$\sqrt{s_{ee}}=500$ GeV, $x=4.8$ for "realistic" spectra.}

\begin{figure}[hbt]
\begin{center}%\vspace{-7mm}
\includegraphics[height=7cm,width=7cm]{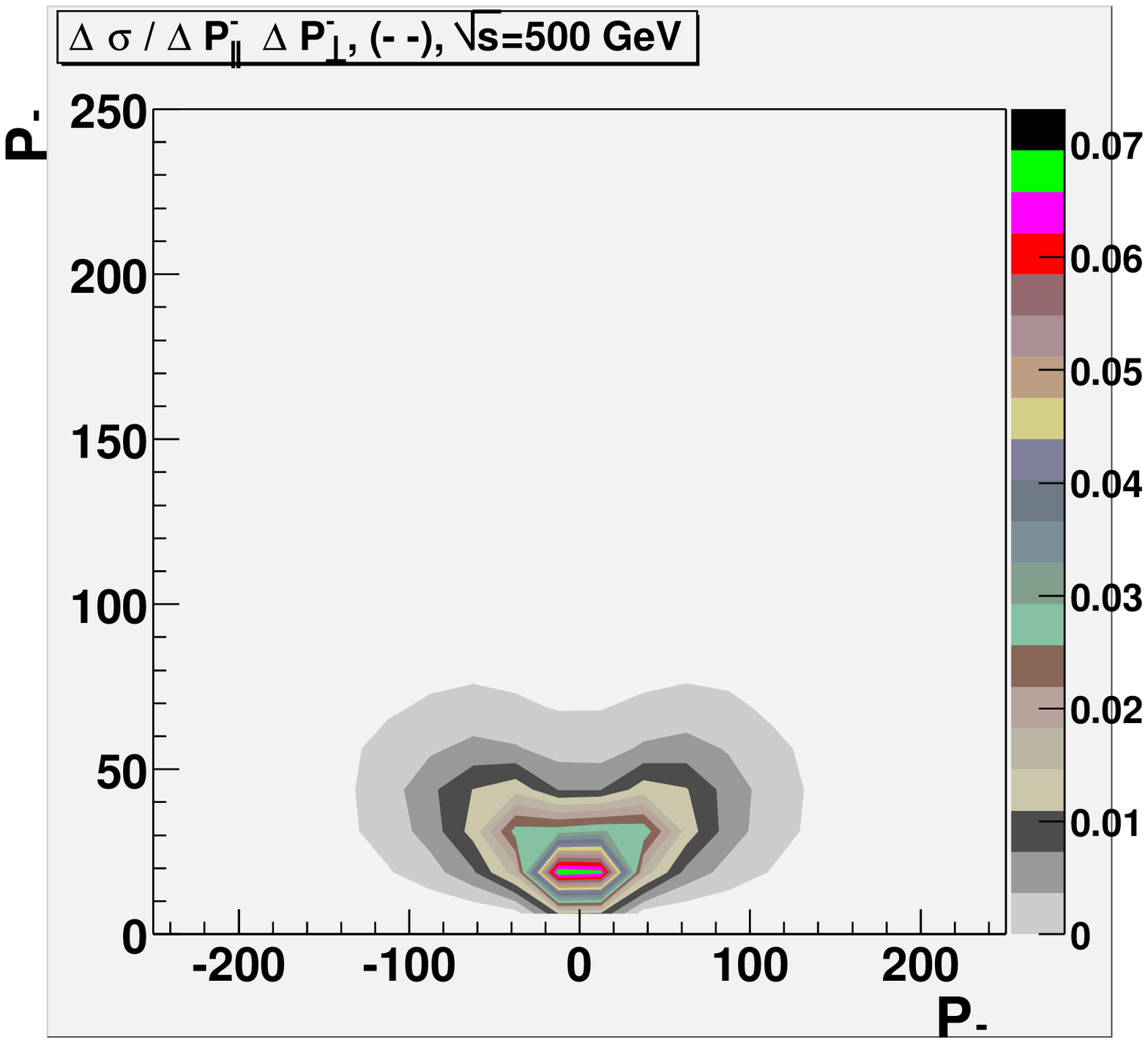}
\includegraphics[height=7cm,width=7cm]{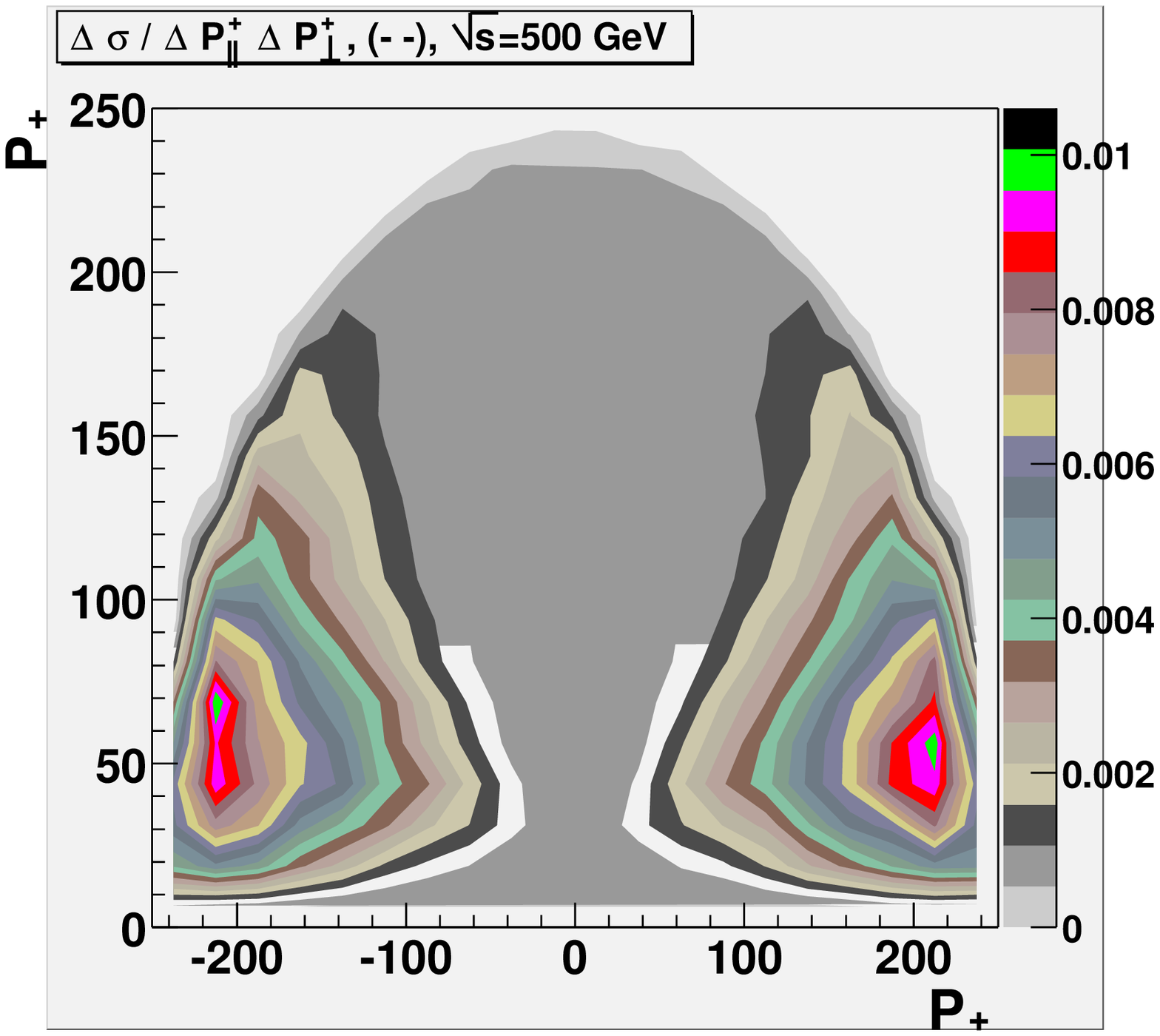}
\includegraphics[height=7cm,width=7cm]{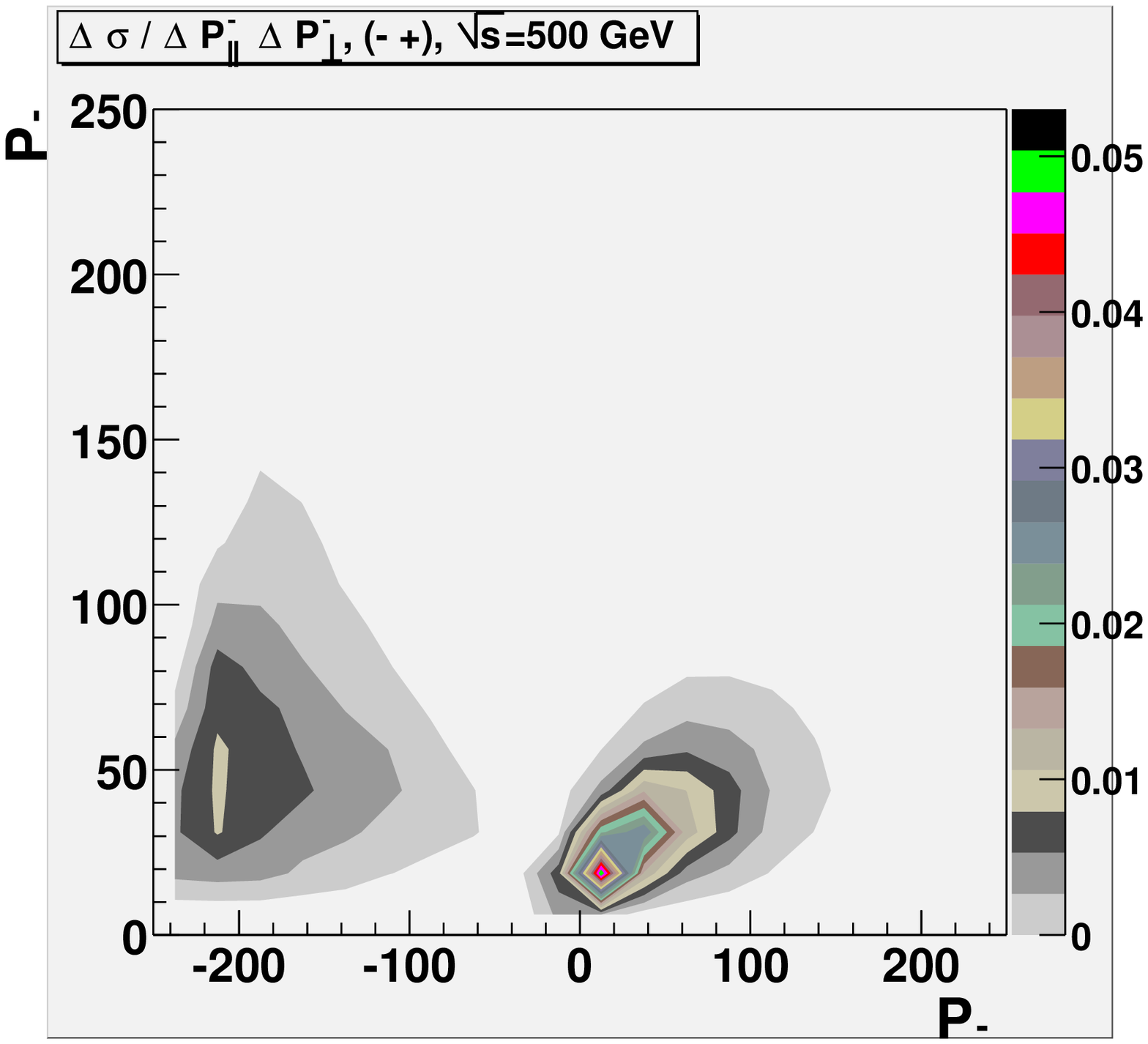}
\includegraphics[height=7cm,width=7cm]{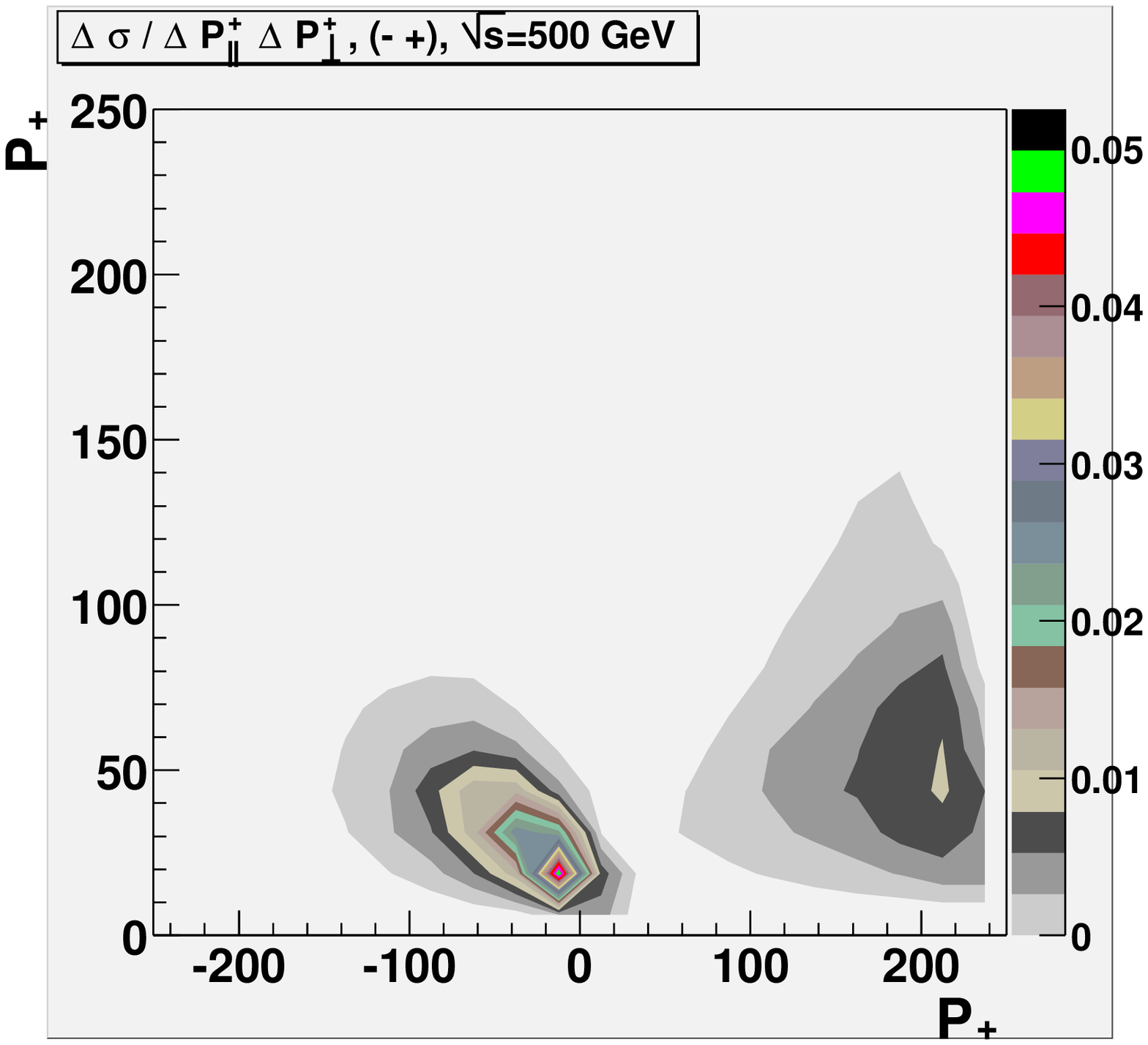}
%\vspace{-7mm} 
\caption{\sf Distributions in the
$(p_\parallel,p_\perp)$ plane (pb/bin). Left -- $\mu^-$, right --
$\mu^+$; from the top: $(--)$,  $(-+)$ -- helicities of colliding
photons. Monochromatic beams. } \label{mm}
\end{center}
%\vspace{-10mm}
\end{figure}

In Fig.~\ref{mm} we show two-dimensional level lines, for
different initial helicity states 
%GeV 
and monochromatic photons.
Note that the scales of effect in the graphs with $\mu^-$ for
$(-\,-)$ case differ from other cases by a factor of seven. Due to
CP conservation, the $\mu^\pm$ distributions for $(-\,-)$ case
coincide with  $\mu^\mp$ distribution for $(+\,+)$ case. The
obtained distributions in the $(p_\|,p_\bot)$ plane have the form
which corresponds to the qualitative picture outlined in
Sec.~\ref{Sec2}.

To make a more transparent analysis, we consider the normalized
mean values of longitudinal and  transverse momenta of $\mu^-$ and
$\mu^+$, determined considering the longitudinal momentum of $\mu$
in the forward hemisphere ($ p_\|>0$), and take their relative
difference as a measure of the charge asymmetry:
 \bea
&P_{L+}^\pm=\frac{\displaystyle\int p_\|^\pm
d\sigma}{\displaystyle E_{\gamma max}\int d\sigma} \,,\quad
P_{T+}^\pm= \frac{\displaystyle\int p_\bot^\pm
d\sigma}{\displaystyle E_{\gamma max}\int
d\sigma}\,,&\label{PLT}\cr & \Delta_L=\frac{\displaystyle P_{L+}^-
-P_{L+}^+}{\displaystyle P_{L+}^- +P_{L+}^+}\,,\quad
 \Delta_T=\frac{\displaystyle P_{T+}^- -P_{T+}^+}{\displaystyle P_{T+}^- +P_{T+}^+}\,.&\label{CHALT}
 \eea
We have calculated these quantities for the process $\ggam\to
W\mu\nu$ at $\sqrt{s}=500$ GeV $\Rightarrow E_{\gamma max}=250$
GeV for monochromatic case and for $\sqrt{s_{ee}}=500$ GeV
$\Rightarrow E_{\gamma max}=207$ GeV for the "realistic" photon
spectra. In tables below the parameter N takes two values N=L
(longitudinal) and N=T (transverse). Table~1 presents the values
of the asymmetries for different photon polarizations in the
monochromatic case. The $(+\,+)$ case coincides with the $(-\,-)$
one with the exchange $\mu^+\to \mu^-$. For $(+\,-)$ initial state
the pictures in forward and backward hemispheres are different.
The latter coincide with that for the forward hemisphere for
$(-\,+)$ initial state with the exchange $\mu^+\leftrightarrow
\mu^-$.
\begin{table}[t]
\begin{center}
\caption {{\sf SM. Monochromatic case. }}
\begin{tabular}{|r|c|c|c|}\hline
$(\lambda_1\,\lambda_2)$\ N & $P_{N+}^-$ & $P_{N+}^+$ & $\Delta_N$ \cr
\hline
 $(-\,-)$\ N=L  & 0.605 & 0.191 & 0.520\cr
N=T & 0.353 & 0.160 & 0.374\cr \hline
$(+\,-)$\ N=L & 0.229 & 0.571 & -0.427\cr
N=T & 0.167 & 0.262 & -0.221\cr \hline
$(-\,+)$\ N=L &0.233 & 0.631 & -0.461\cr
N=T & 0.179 & 0.283 & -0.223\cr \hline
\end{tabular}
%\vspace{-7mm}
\end{center}\label{tabmm500}
%\vspace{-7mm}
\end{table}

One can see that the values of the asymmetry is typically 20-50\%.
That is {\it a huge and easily observable effect.}

Note that the scale of longitudinal distributions is given by
the initial photon energy while the scale for transverse distributions
is given by the $W$ mass. Thus, the transverse asymmetries are
smaller than the longitudinal ones.

$\bullet$ { \bf Effect of photon non--mono\-chro\-ma\-ti\-ci\-ty.}
Photon beams at the Photon collider will be non-monochromatic. The
\begin{table}[b]
%\vspace{-8mm}
\begin{center}
\caption{{\sf SM.}}
\begin{tabular}{|r|c|c|c|}\hline $(\lambda_1\,\lambda_2)$\
N&$P_{N+}^-$&$P_{N+}^+$&$\Delta_N$
\cr \hline
 $(-\,-)$\ N=L  &0.499&0.150&0.536\cr
N=T &0.382&0.178&0.364\cr \hline
$(+\,-)$\ N=L &0.187&0.451&-0.415\cr
N=T& 0.198&0.272&-0.160\cr \hline
$(-\,+)$\ N=L &0.205&0.514&-0.436\cr
N=T&0.219& 0.300&-0.158\cr \hline
\end{tabular}
\end{center}
 \label{tabss500}
\end{table}
energy spectra are well known for high energies where the
approximation of Ref.~\cite{GinK} works well but the low-energy
part will differ depending on technological details. Fortunately,
these photons are unpolarized and do not contribute to the
asymmetry. So that, in all Tables~2--4 we used  "realistic" photon
spectra with $\rho=1$ from Ref.~\cite{GinK} for the high energy part,
 and from Ref.~\cite{Ginzburg1} with $\rho=0$ for the low energy part (similar
to that it was done in Ref.~\cite{Anipko}). We assume here $x=4.8$ for
$e\to \gamma$ conversion which corresponds to the maximal value of
photon energy $E_\gamma^m\approx 207$ GeV at  $\sqrt{s_{ee}}=500$
GeV. The resulting asymmetries are shown in the Table~2. The
longitudinal asymmetry changes only weakly in comparison with the
monochromatic case, while the transverse asymmetry changes
strongly. 
We verified that in the $(p_\|,\,p_\bot)$ plane, relative to the monochromatic case, 
one can see a noticeable deviation in the phase space regions responsible for the asymmetry.

$\bullet$ Next, we consider the effect of { \bf\boldmath  possible
anomalous interactions of $W$--bosons}, parameterized by the usual
anomalous magnetic moment $\Delta\kappa$ and the quadrupole moment
$\lambda$ in the trilinear and quartic vertices (with the same
\begin{table}[h]
%\vspace{-10mm} 
\caption{{\sf SM with anomalous gauge
boson interaction }}
\begin{tabular}{|c|c|r|c|c|c|}\hline
$\Delta\kappa$&$\lambda$&$(\lambda_1\,\lambda_2)$\
N&$P_{N+}^-$&$P_{N+}^+$&$\Delta_N$
\cr \hline
 &&$(-\,-)$\ L  &0.498&0.150&0.537\cr
&&T &0.379&0.182&0.351\cr \cline{3-6}
0.1&0&$(+\,-)$\ L &0.186&0.451&-0.416\cr
& &T & 0.200&0.272&-0.152\cr \cline{3-6}
 &&$(-\,+)$\ L  &0.204&0.510&-0.429\cr
&&T &0.221&0.300&-0.152\cr\hline\hline
 &&$(-\,-)$\ L  &0.491&0.151&0.457\cr
&&T &0.388&0.177&0.378\cr \cline{3-6}
0&0.1&$(+\,-)$\ L &0.190&0.457&-0.413\cr
&&T& 0.192&0.270&-0.169\cr \cline{3-6}
&&$(-\,+)$\ L &0.206&0.522&-0.438\cr
&&T&0.212& 0.301&-0.173\cr \hline
\end{tabular}
\label{tabano500}
\end{table}
% \vspace{-10mm}
normalization as in Ref.~\cite{Anipko}). We calculated the asymmetries
defined in Eq.~(1) for large enough values of $\Delta\kappa$ and
$\lambda$. The results are shown in the Table~3. Comparing with
Table~2, one can see that the anomalous magnetic moment $\Delta
\kappa$ changes the quantities $\Delta_{L,T}$ only weakly. One
should study, whether any other variable is better to investigate
the effects of this anomaly. On the other hand, the quadrupole
moment $\lambda$ changes $\Delta_L$ significantly at least in the
$(-\,-)$ case. Therefore, this asymmetry can be useful for the
study of the $\lambda$ anomaly.

$\bullet$ {\bf Possible new particles.} To mimic the effect of new
interactions and/or new particles, we consider a toy model with a
"muon" having a mass of 40 GeV. The results are shown in Table~4.
Comparing with Table~2, we conclude that the study of charge
asymmetry can be a useful tool for the discovery of new particles.

\begin{table}[t]
\begin{center}
\caption {{\sf Toy model with $m_\mu=40$ GeV}}
\begin{tabular}{|r|c|c|c|}\hline
$(\lambda_1\,\lambda_2)$\ N&$P_{N+}^-$&$P_{N+}^+$&$\Delta_N$
\cr \hline
$(-\,-)$\ N=L  & 0.519 & 0.287 & 0.289\cr
N=T &r 0.385&0.236&0.240\cr \hline
$(+\,-)$\ N=L &0.325&0.569&-0.271\cr
N=T& 0.222&0.310&-0.201\cr \hline
$(-\,+)$\ N=L &0.295&0.545&-0.250\cr
N=T& 0.216&0.308&-0.175\cr \hline
\end{tabular}
\end{center}
\label{tabmuon}
\end{table}
%\vspace{-10mm}

\section{CHARGE ASYMMETRY OF MUONS IN EACH EVENT, \bf\boldmath{$\gamma \gamma \to \mu^+ \mu^- \nu \bar{\nu}$}}

The charge asymmetry in relative distributions of positive and
negative muons can be a more useful instrument to hunt for New
Physics. The first problem here is to find some representative
variables in the 5--dimensional $(\vec{p}_+,\vec{p}_-)$ phase
space. For this purpose we consider the $\gamma \gamma
\to \mu^+ \mu^- \nu \bar{\nu}$ process with monochromatic photons.
\begin{figure}[hbt]
\begin{center}
%\vspace{-9mm}
\includegraphics[height=7cm,width=7cm]{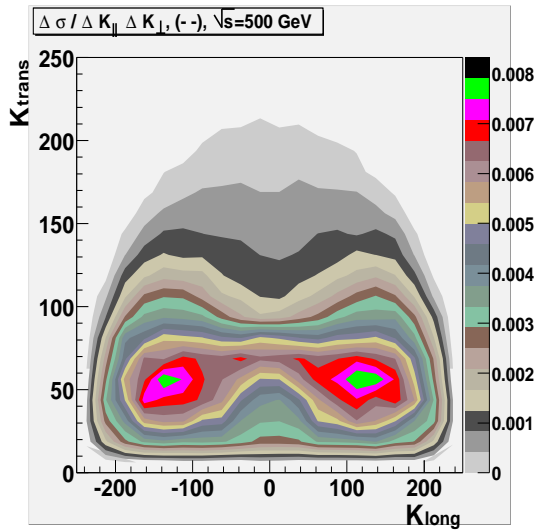}
%\vspace{-7mm}
\caption{\sf Distribution in $k_\|,k_\bot$, monochromatic
photons,} 
\label{kdistr}
\end{center}
%\vspace{-12mm}
 \end{figure}
Fig.~\ref{kdistr} presents the level curves of the distribution in
$\vec{k}=\vec{p}_+ + \vec{p}_-$, in its longitudinal and
transverse component. In the { \em absence } of charge asymmetry
this distribution would be centered around the point
$(k_\|,k_\bot)=(0,0)$. Clearly the charge asymmetry is quite
large.
\begin{figure}[t]
\begin{center}
%\vspace{-7mm}
\includegraphics[height=7cm,width=7cm]{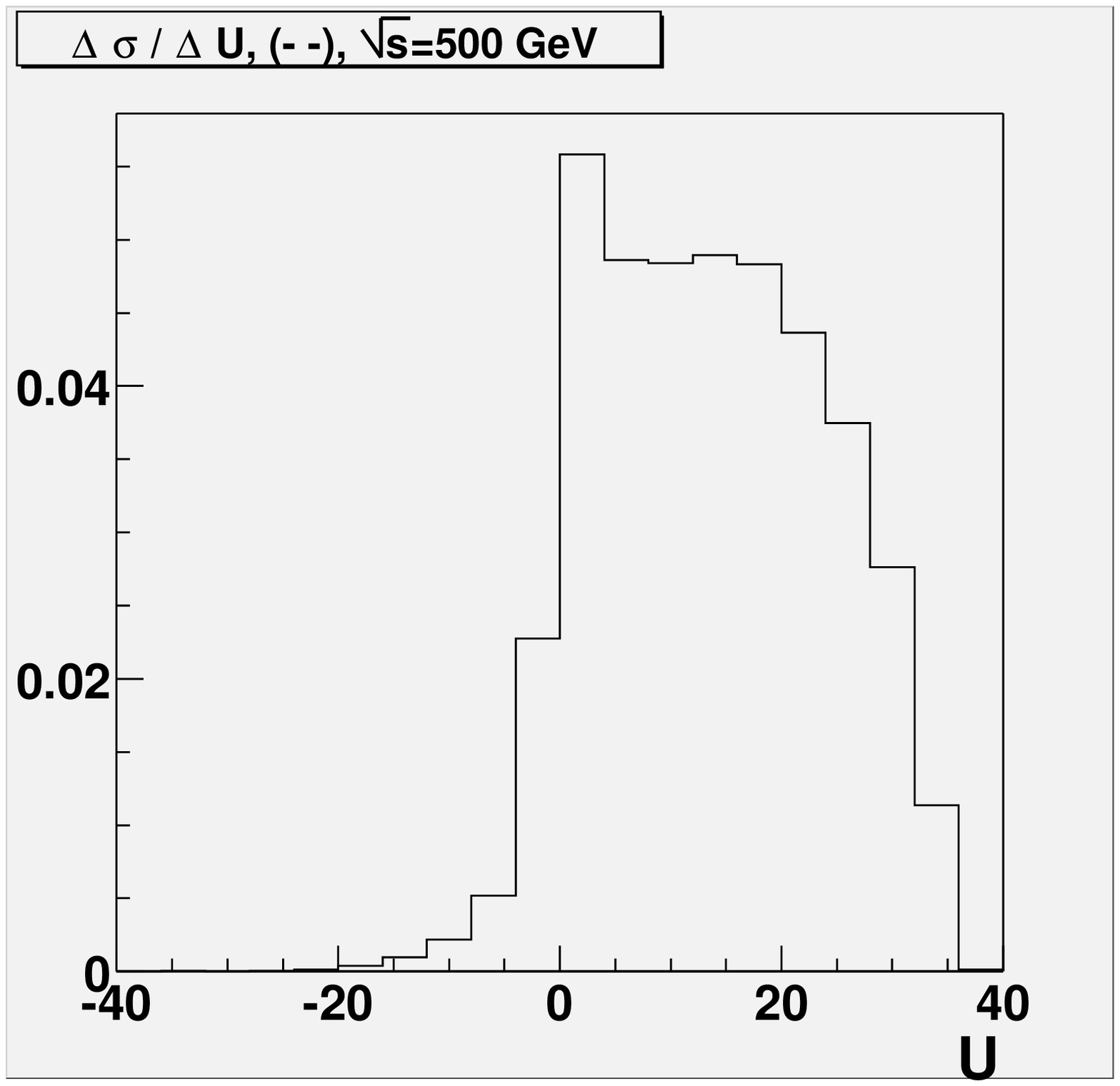}
\includegraphics[height=7cm,width=7cm]{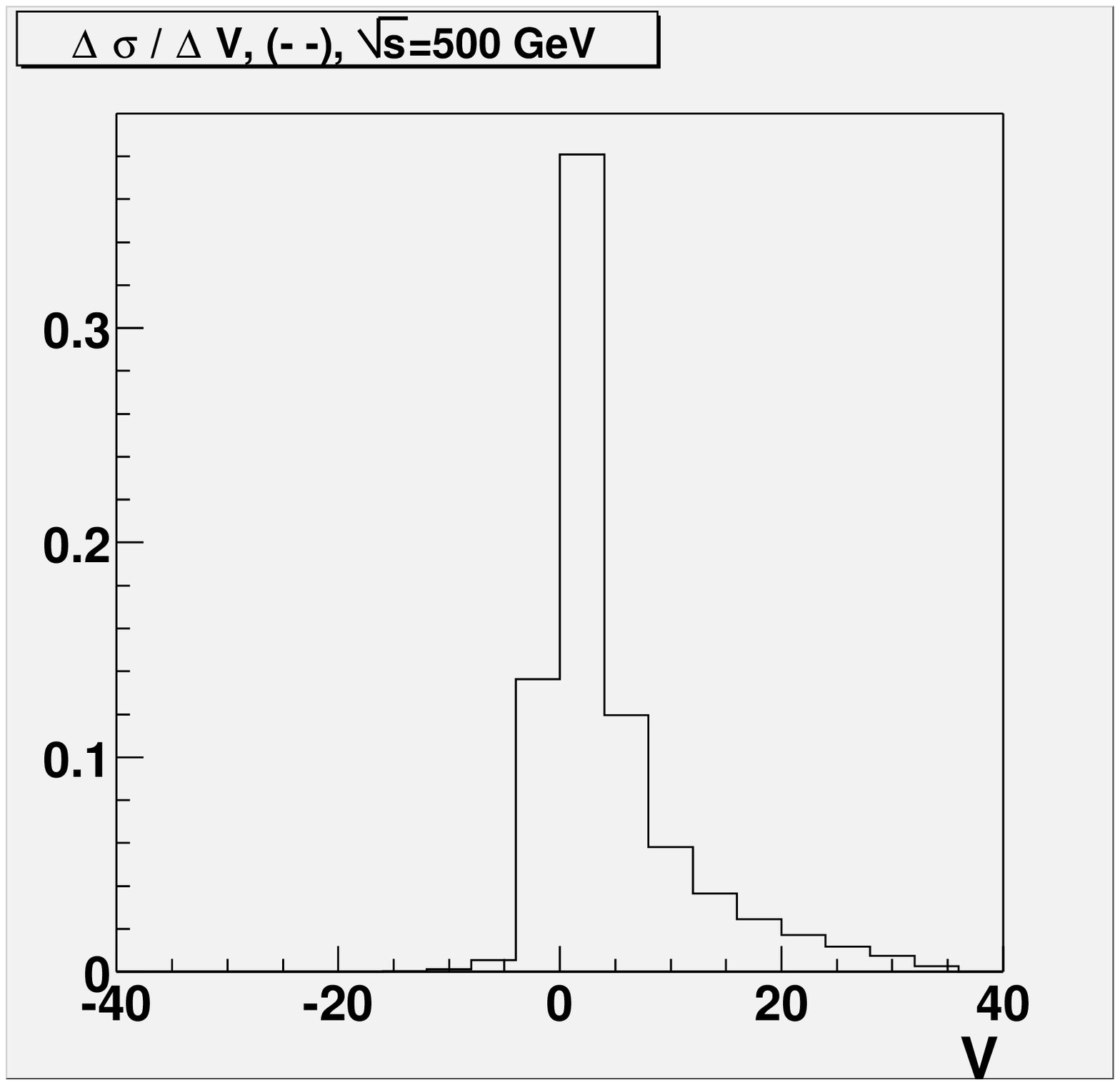}
\includegraphics[height=7cm,width=7cm]{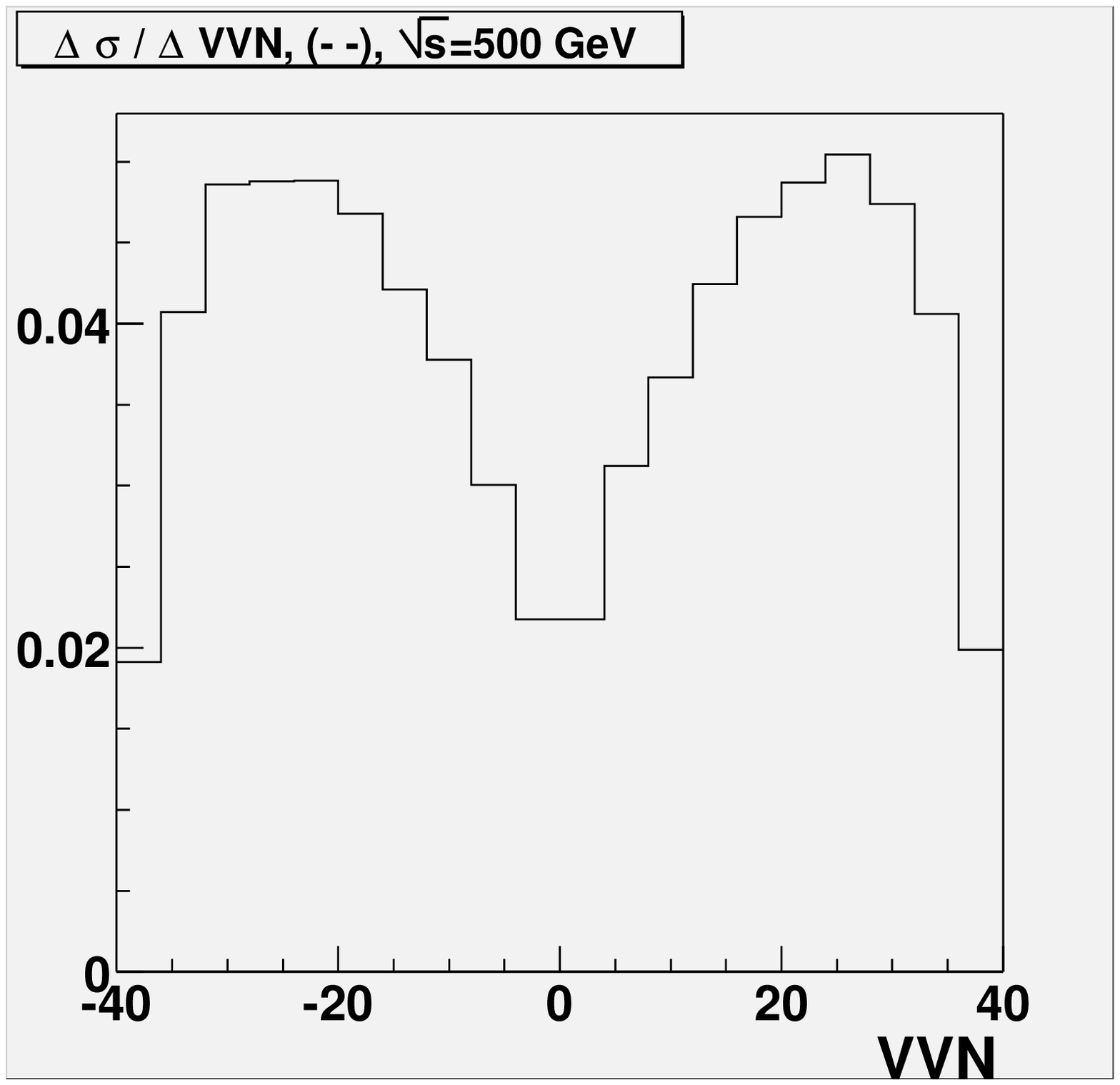}
\includegraphics[height=7cm,width=7cm]{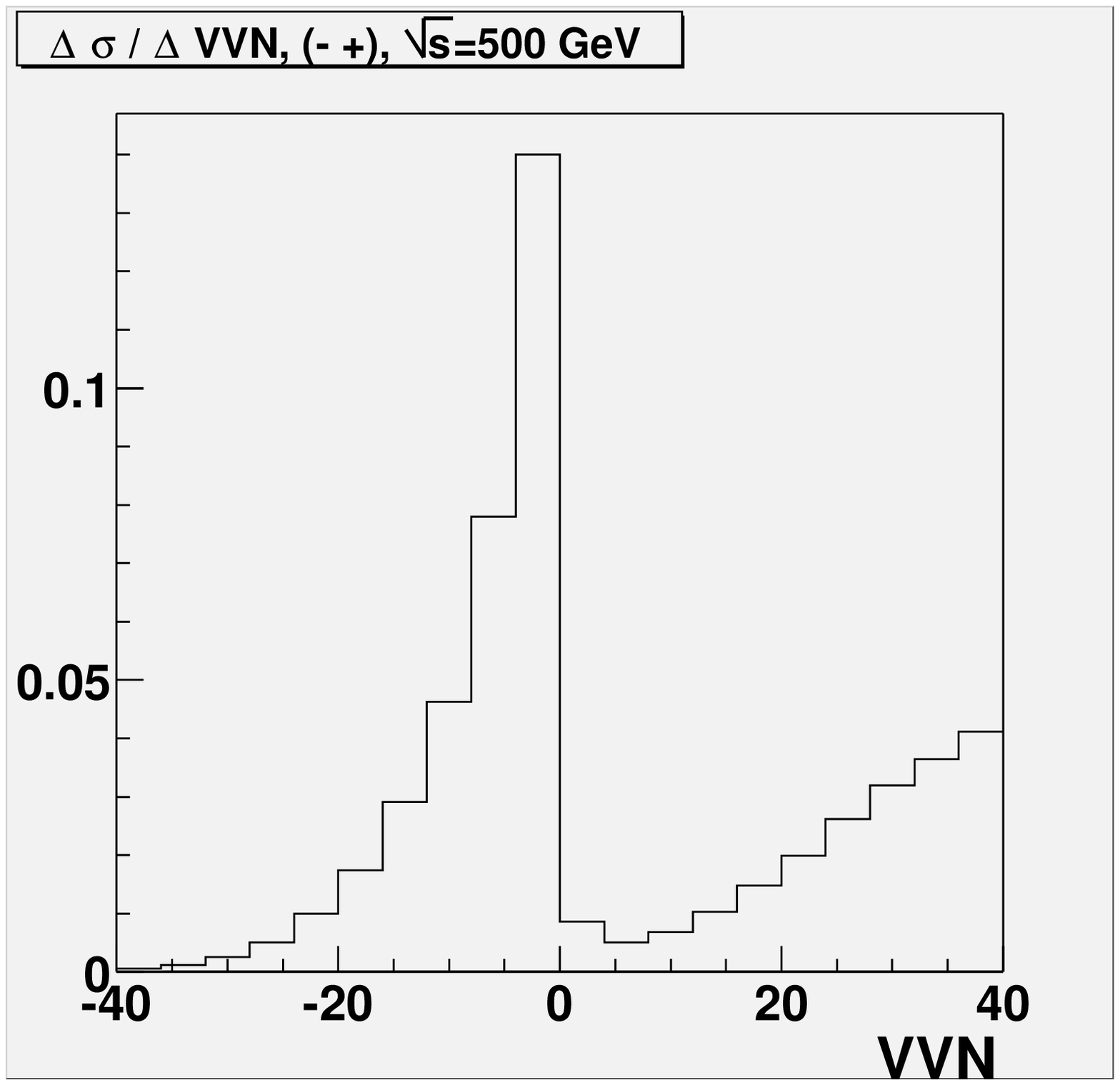}
%\vspace{-7mm}
\caption{\sf Top panels: distribution in $u$ and $v$
for the (- -) case. Bottom panels: distribution in $vvn$ for (- -)
and (-+) initial helicities.} \label{vvn}
\end{center}
%\vspace{-10mm}
\end{figure}
Besides, we study various ``natural'' dimensionless variables. In
particular, we find useful to present distributions in following
variables, Figure~\ref{vvn}.
 \be
\begin{array}{c} v=\frac{\displaystyle 4(p_{\bot
+}^2-p_{\bot-}^2)}{\displaystyle M_W^2},\qquad
u=\frac{\displaystyle 4(p_{\parallel+}^2-p_{\parallel-}^2)}{\displaystyle M_W^2},\\
vvn=\frac{\displaystyle  4(p_{\parallel+}\epsilon_+ -
p_{\parallel-}\epsilon_-)}{\displaystyle  M_W^2}.
\end{array}
 \ee
Fig.~\ref{mm} shows that in the cases $(-\,-)$, $(+\,+)$
distributions in the forward and backward hemispheres are
identical. Therefore, in this case the variables $u$ and $v$ are
useful while $vvn$ can not be useful. Vice versa, for the $(+\,-)$
case the variable $vvn$ is useful while $u$ and $v$ are not.
%can not beuseful. 
The choice of some other variables and sensitivity to New
Physics effects will be studied elsewhere.

\section{DISCUSSION AND OUTLOOK}

$\bullet$ The effects considered so far are identical for
electrons and muons. So that, absolutely the same asymmetry will
be observed in $e^+\,e^-$, $e^+\,\mu^-$, $\mu^+\,e^-$
distributions. Therefore, all these contributions should be
gathered for a complete analysis. This will enhance the value of
cross section for $\gamma\gamma\to\mu^+\mu^-\nu\bar{\nu}$ from 0.9
to 3.7 pb and for $\gamma\gamma\to W^+\mu^-\bar{\nu}$, etc. to
23.5 pb (millions of events per year).

$\bullet$ The contribution to the observable final state is given
not only by the process $\gamma\gamma\to\mu^+\mu^-\nu\bar{\nu}$,
but also by processes $\gamma\gamma\to\mu^+\mu^-\nu\bar{\nu} + \nu\bar{\nu}$ pairs.
The most important of them will be, for example, $\gamma\gamma\to\tau^+\mu^-
\nu\bar{\nu}$, etc. with subsequent decay $\tau\to \mu
\nu\bar{\nu}$. The cross section of this process is 17\% of those
discussed above ($Br(\tau\to \mu \nu\bar{\nu})$ \cite{PDG}) + 17\%
for the case with the change $\tau^+\to \tau^-$, etc. +3\% for
$\gamma\gamma\to\tau^+\tau^-\nu\bar{\nu}$. Unfortunately, at the
moment there are no regular methods for precise calculation of
such processes like CompHEP or WHIZARD (6 or 8 particles in final
state and very narrow intermediate state). We hope that the
corresponding algorithm will be developed in a reasonable time
using the fact that $\tau$ is a very narrow particle.

$\bullet$ We plan to calculate such a charge asymmetry in the
case of existence of some new particles and interactions
(e.g. MSSM).

\begin{acknowledgments}

This work is partially supported by grants RFBR 02-02-17884,
NSh-2339.2003.2, INTAS 00-00679 and grant 015.02.01.16 Russian
Universities and  by the European Union under contract N.
HPMF-CT-2000-00752.
I.~F.~Ginzburg acknowledges support from the Landau Network -- Centro Volta
(Como) Italy, and from INFN Sezione di Perugia.
\end{acknowledgments}


\begin{thebibliography}{99}

\bibitem{Ginzburg1}
I.~F.~Ginzburg, G.~L.~Kotkin, V.~G.~Serbo and V.~I.~Telnov,
%``Colliding Gamma E And Gamma Gamma Beams Based On The Single Pass Accelerators (Of Vlepp Type),''
Nucl.\ Instrum.\ Meth.\   205 (1983) 47;\\
I.~F.~Ginzburg,
G.~L.~Kotkin, S.~L.~Panfil, V.~G.~Serbo and V.~I.~Telnov,
%``Colliding Gamma E And Gamma Gamma Beams Based On The Single Pass E+ E- Accelerators. 2.
%Polarization Effects. Monochromatization
%Improvement,''
Nucl.\ Instrum.\ Meth.\ A  219 (1984) 5.
%%CITATION = NUIMA,A219,5;%%

\bibitem{Tesla}
B.~Badelek {\it et al.}  [ECFA/DESY Photon Collider Working Group
                  Collaboration],
%``TESLA Technical Design Report, Part VI, Chapter 1: Photon collider  at TESLA,''
hep-ex/0108012.
%%CITATION = HEP-EX 0108012;%%

\bibitem{GKPS} I.F. Ginzburg, G.L. Kotkin, S.L. Panfil, V.G. Serbo,
Nucl.\ Phys.\ B 228 (1983) 285.

\bibitem{Pukhov}
A.~Pukhov {\it et al.},
%``CompHEP: A package for evaluation of Feynman diagrams and integration  over multi-particle phase
%space. User's manual for version 33,'',
hep-ph/9908288.
%%CITATION = HEP-PH 9908288;%%

%\cite{Boos:1997jt}
\bibitem{Boos}
E.~Boos and T.~Ohl,
%``Towards a complete calculation of gamma gamma $\to$ 4f,''
Phys.\ Lett.\ B 407 (1997) 161, hep-ph/9705374.
%%CITATION = HEP-PH 9705374;%%

\bibitem{BBB} M.Baillargeon, G. Belanger, F. Boudjema.
hep-ph/9405359

\bibitem{GinK} I.F. Ginzburg, G.L. Kotkin,
Eur.\ Phys.\ J.\ C 13 (2000) 295.

\bibitem{Anipko}
D.~A.~Anipko, I.~F.~Ginzburg and A.~V.~Pak,
%``The gauge boson anomalous interactions via process e-  gamma $\to$ W- nu. The lepton decay mode,'',
hep-ph/0201072.
%%CITATION = HEP-PH 0201072;%%

\bibitem{PDG}
K.~Hagiwara et al.,
Phys.\ Rev.\ D 66, 010001 (2002)


\end{thebibliography}
\end{document}